# Localized Radiofrequency Heating for Enhanced Thermoelectric Energy Generation Using Natural Galena Ore

Karthik R[1], Yiwen Zheng[2], Rajesh Kumar Sahu[3,4], Punathil Raman Sreeram[1*], Soma Banik[3,4], Aniruddh Vashisth[2*], Chandra Sekhar Tiwary[1*]

[1]*Department of Metallurgical and Materials Engineering, Indian Institute of Technology Kharagpur, West Bengal, India*

[2]*Department of Mechanical Engineering, University of Washington, Seattle, WA, USA*

[3]*Accelerator Physics and Synchrotrons Utilization Division, Raja Ramanna Centre for Advanced Technology, Indore, 452013, India.*

[4]*Homi Bhabha National Institute, Training School Complex, Anushakti Nagar, Mumbai, 400094, India.*

*Corresponding email: sreerampunam@metal.iitkgp.ac.in (PRS), somasharath@gmail.com (SB), vashisth@uw.edu (AV), chandra.tiwary@metal.iitkgp.ac.in (CST)*

## Abstract

The efficiency of thermoelectric devices can be significantly enhanced by maintaining a stable temperature gradient, which can be achieved through localized heating. Radio waves serve as an ideal heat source for this purpose. In this study, we demonstrate the enhancement of thermoelectric performance in earth-abundant natural ore Galena (PbS) through localized radio frequency (RF) heating. RF heating experiments conducted at frequencies between 35 and 45 MHz induced substantial localized heating in PbS, generating a temperature gradient of 32 K. This resulted in a Seebeck voltage of -5.8 mV/K, approximately 13 times greater than the conventional Seebeck coefficient of PbS (440 µV/K). Additionally, a power factor of 151 mWm$^{-1}$K$^{-2}$ and an overall RF to thermoelectric conversion efficiency of 15% were achieved. Molecular dynamics simulations corroborate the experimental findings, providing insights into the mechanism of thermal transport and RF induced heating in PbS. These results highlight the potential of localized RF heating as an effective strategy for enhancing thermoelectric performance, with promising implications for ambient thermoelectric energy harvesting applications.

# Introduction

With the rise in demand for sustainable and innovative energy harvesting methods, thermoelectricity has gained attention due to its ability to convert waste heat to electricity. Thermoelectric generators (TEG) are particularly well suited for operation in extreme environments, including isolated regions of earth, space and planetary surfaces. The efficiency of a TEG is governed by the interplay of three key material properties: the Seebeck coefficient (S), electrical conductivity (σ), and thermal conductivity ($\kappa$). Specifically, efficiency scales proportionally with both Seebeck coefficient and electrical conductivity while being inversely related to thermal conductivity. However, optimizing these parameters presents a fundamental challenge—materials with a high Seebeck coefficient often exhibit lower electrical conductivity, and vice versa. Additionally, minimizing thermal conductivity is crucial, as it helps sustain the temperature gradient essential for efficient thermoelectric performance. A stable temperature gradient is essential for the continuous generation of thermoelectric voltage in a TEG. This relationship is quantitatively described by the Seebeck coefficient, given as: S = dV/dT, where S represents the Seebeck coefficient, dV is the differential change in voltage, and dT is the corresponding temperature difference. This principle highlights the necessity of maintaining a stable thermal gradient to drive charge carrier diffusion and sustain electrical power output. Thermal conductivity ($\kappa$)in thermoelectric materials consists of primarily two components: electronic thermal conductivity ($\kappa_e$) and lattice thermal conductivity ($\kappa_l$). Expressed as $\kappa = \kappa_e + \kappa_l$. Since electronic thermal conductivity is intrinsically linked to electrical conductivity via the Wiedemann-Franz law, reducing $\kappa_e$ can negatively impact charge transport. Therefore, achieving high thermoelectric efficiency requires strategies that minimize lattice thermal conductivity ($\kappa l$) while preserving electrical conductivity, striking an optimal balance between these competing factors. The lattice thermal conductivity can be varied through the introduction of point defects [1], dislocations [2], herringbone structures [3] and nanostructuring[4] techniques. However, the consequence of lowering thermal conductivity using the above approach is that it may reduce the material's overall electrical conductivity, resulting in a low Seebeck coefficient. Several other physical approaches like lowering carrier concentration[5–7], energy filtering[8,9], quantum confinement effect[10–12], Spin caloritronic effect[13–16] and Phonon drag effect[17,18] have been explored to enhance the Seebeck coefficient. A unique approach is to use localised heat sources to generate a thermal gradient without altering the chemistry of materials. Researchers have demonstrated this approach by utilizing a pulsed LASER source to heat materials and create a temperature gradient. This has led to the development of Photo thermoelectric effect (PTE) based photodetectors using 2D Te and $MoS_2$ monolayer in which the photogenerated hot carriers are utilized for photoenergy conversion through thermoelectric effect, thereby enhancing the sensitivity of photodetectors[19–21]. Beyond material engineering, efficiency can also be enhanced by optimizing heat management strategies, ensuring effective thermal gradients across the device by generating the heat from electromagnetic radiations available in the ambient, such as radio waves and microwaves. This can be integrated with thermoelectric materials to harvest ambient electromagnetic energy and utilise it for charging devices.

In this work, we present an innovative approach to harness RF (Radiofrequency) energy by utilizing metal-semiconductor junctions (Schottky junctions) as localized heating sources to for thermoelectric energy generation. We demonstrate that these junctions effectively generate localised heat, creating a significant thermal gradient that enhances thermoelectric power generation. **Figure 1** provides a comparative overview of various strategies for improving the Seebeck coefficient (Thermopower), highlighting the advantages of our approach over conventional methods. To validate our concept, we utilize natural ore Galena (PbS), an earth abundant, n-type semiconductor with a narrow bandgap of approximately 0.4 eV, making it highly suitable for optoelectronic, Radio frequency detection (RF) and energy applications. The strong light absorption in near-and mid-infrared regions enables its widespread use in infrared photodetectors, thermal imaging, and night vision systems[22]. Additionally, PbS exhibits low thermal conductivity and a high Seebeck coefficient, making it a promising candidate for thermoelectric energy harvesting [23]. The ease of synthesis, stability and tunability at the nanoscale further enables its integration into modern photovoltaic cells, quantum dots-based optoelectronics, and advanced sensing devices [24,25]. The presence of intrinsic defects and localized states in natural galena enhances carrier transport, allowing it to function as an effective detector for weak radio signals. In our previous work on 2D PbS, we have already shown the application of PbS for Radio frequency energy harvesting applications[26]. In this work, we, for the first time, integrate the radio frequency (RF) detection capability and thermoelectric properties of galena to enable efficient thermoelectric energy generation. We fabricated a Schottky junction on the surface of a cleaved PbS crystal using a titanium (Ti) tip, which serves as an RF detector, inducing localized heating. RF heating studies conducted across various frequencies, followed by thermoelectric voltage measurements, revealed a significant enhancement in the Seebeck voltage. Additionally, we employed Molecular Dynamics (MD) simulations to investigate the thermal transport mechanism in PbS under RF-induced localized heating. Our findings demonstrate that RF driven localized heating in galena can effectively enhance thermoelectric efficiency by generating substantial thermal gradients within the material.

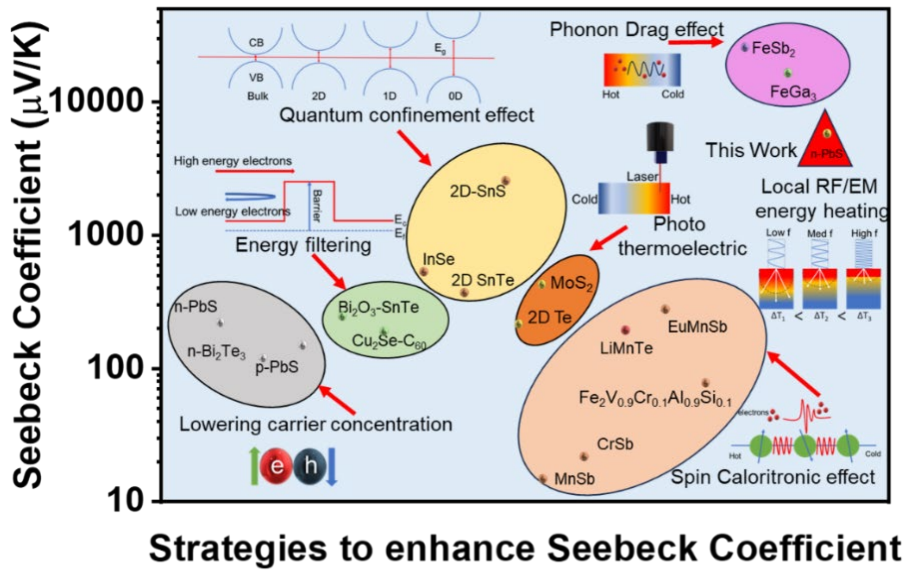

*Figure 1- Schematic representation showing various approaches adopted to enhance the Seebeck coefficient in a variety of thermoelectric materials.*

# Experimental results and discussion

To comprehensively characterize the structural and chemical properties of PbS samples, we conducted various analytical techniques. **Figure 2a** shows the XRD pattern for bulk and polished PbS samples along with the reference pattern for PbS (ICSD-600243). PbS belongs to a cubic crystal system (Fm-3m) with lattice parameter a=b=c= 5.94 Å. In bulk PbS samples, the observed several diffraction planes suggest the polycrystalline nature of the sample. In the case of the polished sample, sharp intense peaks at the (200) and (400) planes are observed indicating the cleavage plane; the maximum orientation of the polished sample is along those planes. However, a peak shift to higher angles is observed in the case of a polished sample which indicates lattice contraction suggesting a sulfur-deficient or lead-rich surface. EBSD studies were also performed to substantiate the orientation plane in the polished sample (**Figure 2b**). The EBSD grain color map image also indicates the maximum orientation of atomic planes is along (001) family of planes. The FESEM image of the polished sample is shown in **Figure**

**2c** revealing a single phase. The EDS spectra of the polished PbS are provided in a supporting information (**Figure S1**) indicates the purity of polished PbS without any significant contamination. The vibrational nature of the polished sample is also analyzed using Raman spectroscopy as shown in **Figure 2d**. Two intense vibrational modes are observed at 216 cm$^{-1}$ and 278 cm$^{-1}$, corresponding to 1 longitudinal optical (LO) mode and two phonon modes, and the weak vibration at 480cm$^{-1}$ indicates 2LO mode in PbS [26]. The chemical stability of the polished surface of PbS is also verified using XPS studies. The XPS survey scan plot is provided in the supporting information (**Figure S2**). XPS binding energy plots for Pb 4f and S 2p are shown in **Figures 2e** and **2f**, respectively. The observed major peaks at 136.8 eV (Pb 4f$_{7/2}$) (Pb$^{2+}$) and 141.6 eV (Pb 4f$_{5/2}$) (Spin-orbit split) exhibit a negative binding energy shift compared to reported Pb 4f values of PbS [27,28]. This shift indicates an increased electron density on Pb atoms suggesting a Pb-rich surface. Similarly, the S 2p spectrum shows a downward peak shift with peaks at 160 eV (S 2p$_{3/2}$) (S$^{2-}$) and 161.15 eV (S 2p$_{1/2}$) (Spin-orbit split), further confirming the Pb-rich nature of the cleaved surface. This observation is consistent with XRD analysis, which also supports the presence of a Pb-rich surface. Additionally, the absence of extra peaks in the XPS spectrum confirms that the sample remains unoxidized, preserving its purity.

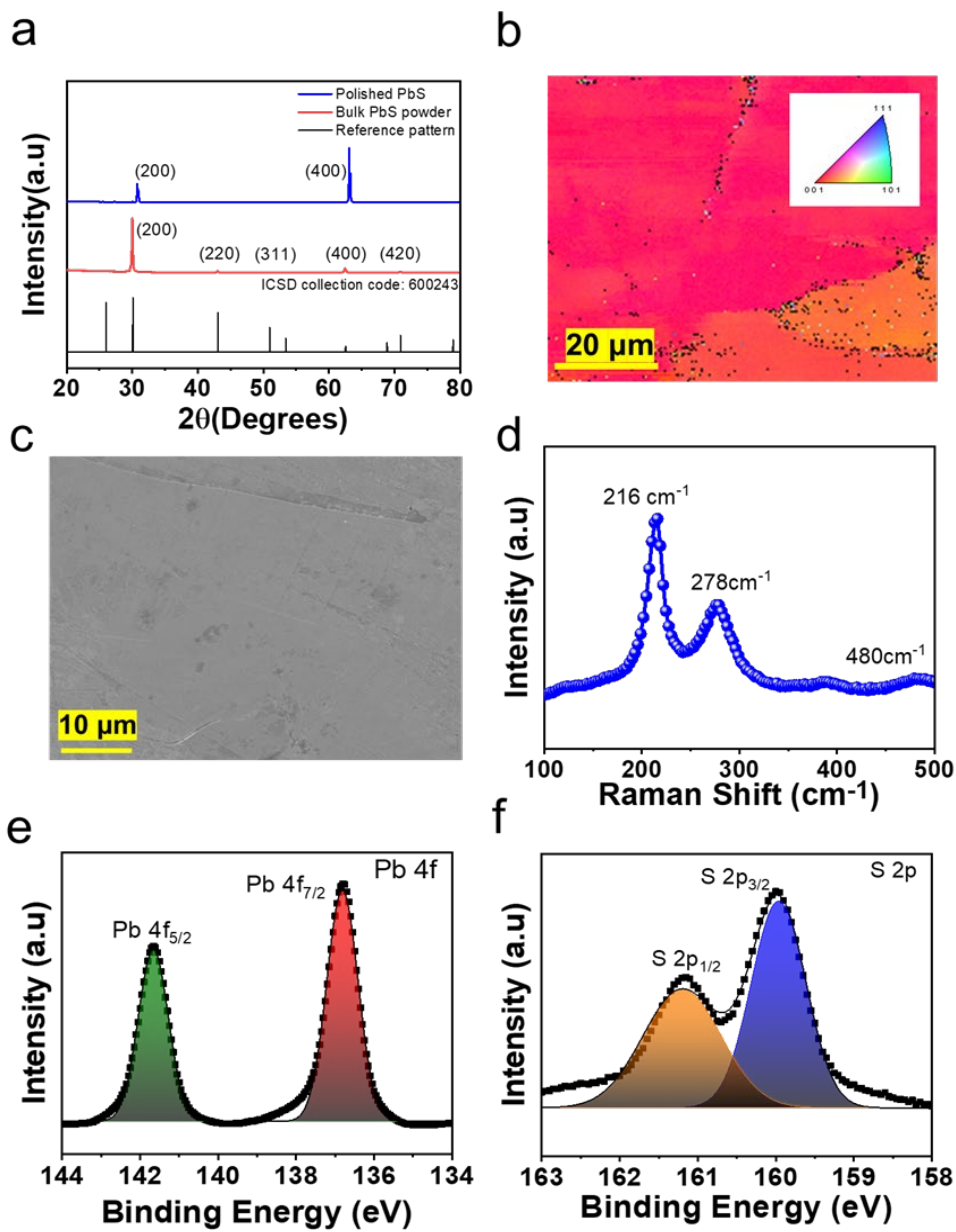

*Figure 2*- (a) XRD pattern of bulk PbS and polished PbS samples, (b) EBSD grain color map of polished PbS sample showing maximum orientation along (001) plane, (c) FESEM image of polished PbS sample, (d) Raman spectra of polished PbS sample, (e, f) XPS spectra of PbS showing separate plots for Pb 4f and S 2p binding energy plots.

## Photoemission Studies on Galena

Temperature-dependent ARPES and valence band measurements were carried out to understand the changes in the band structure for investigating the electronic and thermoelectric properties in the natural ore Galena PbS crystal. PbS is preferentially cleaved along (001) and hence, it is expected to see the square face of the surface Brillouin zone (BZ) as shown in

**Figure 3a**. The Γ and L points are projected on the $\bar{\Gamma}$ and $\bar{X}$ points on the surface BZ while the other symmetry points X, K and W are projected onto the $\bar{M}$ point on the surface BZ. In **Figure 3 (c-f)** we have shown the ARPES band mapping of the PbS surface along the $\bar{X} - \bar{\Gamma} - \bar{M}$ direction at 300 K, 200 K, 100 K and 20 K. At 300 K in **Figure 3c,** we have observed a localized flat band near the Fermi level ($E_F$) at -0.64 eV. The other bands at higher binding energy below -1.5 eV are not well resolved at 300 K due to the thermal broadening of the bands. As the temperature decreases, we find that the band at -0.64 eV becomes more localized and there is an emergence of an electron-like conduction band near the $E_F$ at 200 K (shown by white dots in **Figure 3d**) At 100 K and 20 K, we find that there is a clear signature of an electron-like conduction band and hole-like valence band (marked with the white dots in **Figure 3 e** and **Figure 3f**) at the $\bar{X}$ point. The occurrence of such an electron-like conduction band is well reported in n-type PbS[29]. The intensity of the bulk valence bands was found to increase with the decrease in temperature. A constant energy gap of ≈0.48 eV at $\bar{M}$ point is observed at all temperatures. The coexistence of electron-like conduction bands at $\bar{X}$ point and an energy gap at $\bar{M}$ point (**Figure 3f**) indicates two different natures of valence bands present in this system with different effective masses. These band states are denoted as localised heavy band (HB) with an edge lying at 0.48 eV and light hole band (LB) with the edge lying at 0.33 eV in **Figure 3f**. The schematic E-k diagram portrayed from the ARPES data is shown in **Figure 3b**. The appearance of similar distinct valence band maxima which are separated in energy as well as in momentum are also reported for n-type PbS single crystals[29]. From the ARPES measurement, we have concluded that an increase in temperature increases the thermal excitation in the system which leads to the broadening of the HB and LB that results in the merging of both the bands. This phenomenon is prominently visible at 200 K in **Figure 3d**. Further increase in temperature leads to the thermal excitation of the electrons from HB to the CB.

To understand the change in density of states (DOS) in PbS as a function of temperature we have shown the energy distribution curves (EDC) in supporting information **Figure S3 a**. The inelastic background has been subtracted from the raw data using the Tougaard method [30] and normalized at $E_F$ = 0 eV. There are prominent 5 features observed in the EDCs marked as A, B, C, D and E at -0.1 eV, -0.64 eV, -1.57 eV, -2.64 and -3.2 eV. All the features in the VB arise due to the hybridization between Pb 6s-6p states with the S 3p states. The VB of PbS natural ore shows a very good matching with the DFT calculation of bulk PbS as reported [31–33]. To see the behaviour of HB, LB and CB as a function of temperature we have shown the second derivative of the EDC in **Figure S3 b**. The energy position of the HB, LB and CB are marked at -0.1 eV, -0.33 eV and -0.64 eV, respectively and shown by dotted lines in **Figure S3 b**. We can see that the position of these bands does not change with the increase in temperature but there is a thermal broadening of the bands which results in the merging of the localized HB with the hole-like LB. Thus, there is a spectral distribution of the localized states which leads to an increase in the density of heavy holes in the valence band near to the conduction band. As the temperature is increased, the charge transport in PbS will be dominated by the heavy holes due to the thermal excitation which results in enhanced Seebeck coefficient and higher thermoelectric power factor at higher temperatures. Hence, the changes in the spectral shape of the bands due to the thermal broadening results in superior thermoelectric properties of PbS at higher temperatures.

*Figure 3*-Band mapping of PbS (001) showing (a) the surface Brillouin zone, (b) the schematic E-k diagram portrayed from the ARPES data recorded using 56 eV excitation energy at (c) 300 K, (d) 200 K, (e) 100 K and (f) 20 K.

**Thermoelectric Measurements of PbS**

To assess the thermoelectric properties of PbS we performed thermoelectric measurements on polished PbS samples provided in supporting information. **Figure S 4a** shows the temperature-dependent DC electrical conductivity of PbS. It is observed that temperature increases the conductivity of PbS due to degenerate semiconducting nature. **Figure S4b** shows temperature-dependent thermal conductivity plots of PbS. With the increase in temperature, the thermal conductivity decreases due to enhanced phonon scattering due to thermal excitation. The Seebeck coefficient measurements as shown in **Figure S4c** show an increment of Seebeck coefficient with temperature indicating n-type behaviors of PbS where electrons as the majority carriers. Finally in **Figure S4d**, to assess the impact of Radiofrequency on PbS, we performed AC electrical conductivity in a desired RF range (35-50MHz). We chose this frequency range to avoid possible interference with commercial radio transmissions. The AC electrical conductivity of PbS is very low at low frequencies and it increases with frequency. This is evidence of a strong skin effect in PbS at high frequencies. Since such a low conductivity value is observed, the resistivity and therefore the skin depth value will be significantly high. This means that the low-frequency electric field component of radio waves will be able to penetrate the material to a greater extent that can excite or oscillate more electrons resulting in the generation of thermal electrons. We believe this could result in localized heat generation in materials which can be utilized for thermoelectric energy harvesting.

**Electrical measurements of Ti/PbS junction**

To evaluate the electrical characteristics of the Ti/PbS junction, we set up a detailed experiment, including the device configuration, voltage-current measurements, and analysis of Schottky diode behaviour and capacitance. **Figure 4a** shows a schematic representation of a device setup configuration for measuring electrical properties. The original photograph is provided in the supporting information (**Figure S5**). The setup consists of a polished PbS sample embedded in an epoxy matrix. A copper wire is soldered on one end of the surface to make electrical contact with the polished sample. The other connection is made through a point contact using a Ti tip. To establish the nature of electrical contacts between Ti and PbS, voltage-current characteristics were performed. **Figure 4b** shows the temperature-dependent voltage-current characteristics of the setup. It is observed that at a certain bias voltage, the current increases exponentially which resembles a Schottky diode characteristics. **Figure 4c** shows the forward bias voltage-current characteristics, where the current is increasing with temperature satisfying thermal activation mechanism. **Figure 4d** shows a linear relationship between $\ln(J/T^2)$ and $V^{0.5}$ indicating Schottky emissions at a voltage less than 1V. **Figure 4e** also shows a linear relationship between $\ln(J_0)$ and V further validating the above finding. Schottky barrier height is an important property which says about the rectifying ability of the diode. Before calculating the Schottky barrier height, the effective Richardson constant was obtained from the slope of **Figure 4f**. The temperature-dependent Schottky barrier height of the sample is shown in **Figure 4g** where a maximum barrier height of 0.54 eV is obtained. A low barrier height ensures easy movement of carriers across the barrier at low input voltage. The low electronic bandgap of 0.42 eV as shown in **Figure 4h** also suggests an easy transition of carriers from the valence band to the conduction band at low input voltage or thermal excitations. The capacitance measurement of the Ti/PbS junction is shown in **Figure 4i**, as frequency increases the capacitance of the junction decreases indicating low capacitive impedance at high frequencies. This property of the Schottky junction is suitable for high-frequency applications since the carrier won't face much energy loss at the junction.

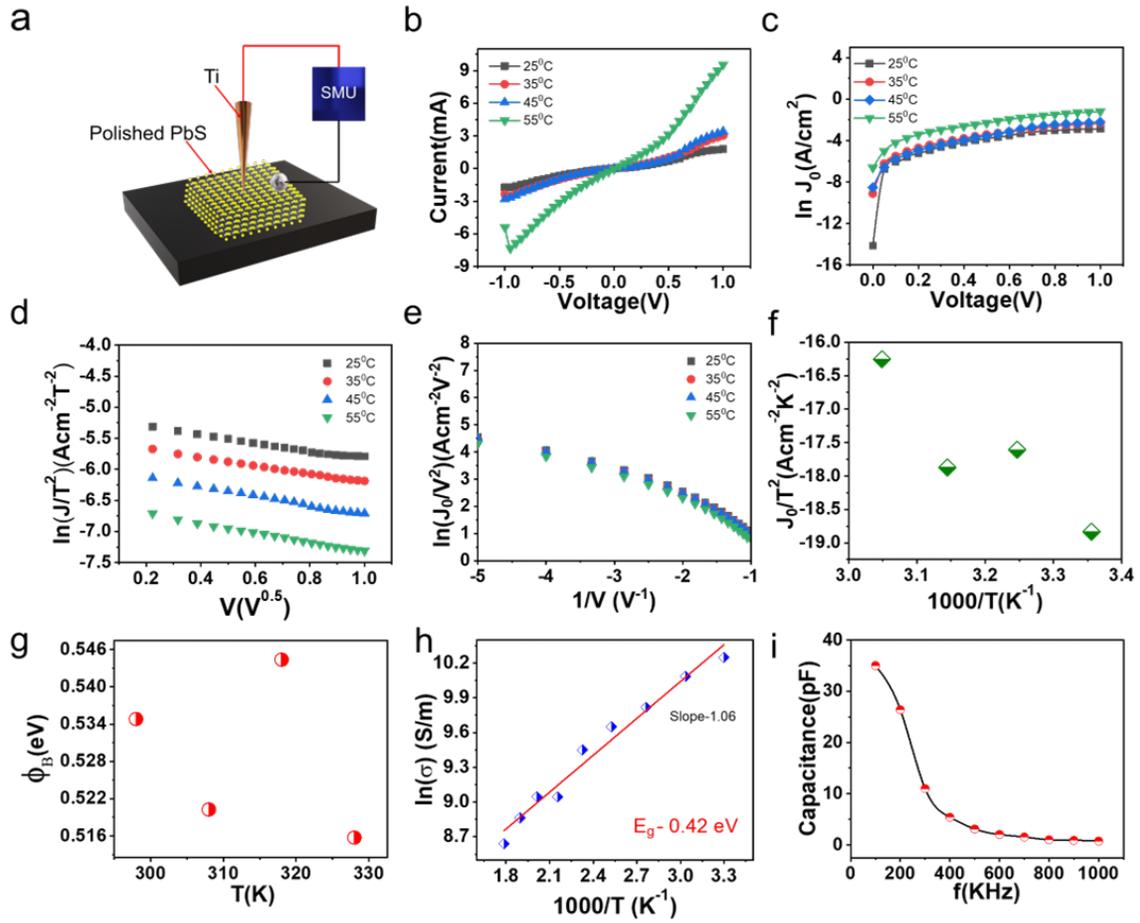

*Figure 4- (a) Photograph of polished PbS sample embedded in epoxy with electrical contacts, (b) Temperature-dependent voltage-current characteristics of Ti/PbS junction, (c) Forward bias characteristics of Ti/PbS junction, (d) ln(J/T$^2$)-V$^{0.5}$ plot for verifying Schottky emission, (e) ln (j$_0$)-V$^{-1}$ plot for verifying Schottky emission, (f) Richardson plot, (g) Temperature-dependent barrier height plot, (h) Arrhenius plot showing electronic bandgap of PbS, (i) Frequency-dependent junction capacitance of Ti/PbS.*

**Thermoelectric studies of Ti/PbS by localized RF heating**

With the confirmation of a strong skin effect in PbS at RF ranges, we engineered a device-like configuration to harness RF and induce local RF heating in PbS. Given that Ti/ PbS interface exhibits a Schottky diode behavior, it has the potential to function as a radio signal detector. In conventional radio receivers, an ideal Schottky diode with strong rectification is used to rectify the incoming RF signal. However, V-I studies reveal significant leakage currents, which could typically lead to poor reification performance. Despite this limitation, we propose that for strong rectification is not a prerequisite for RF heating. Instead, the large leakage currents switching at high frequencies are expected to generate heat at the metal-semiconductor interface which can be effectively utilized for localized heating. To validate this hypothesis, we conducted radiofrequency studies on Ti/ PbS coupled with thermal imaging. **Figure 5(a, b)** presents a schematic representation of the experimental setup, which consists custom-built RF signal generator of maximum power 1W and a lab-fabricated RF thermoelectric system. A photograph of the experimental setup is provided in supporting information (**Figure S6**). For electrical measurements we employed an analogue milli voltmeter, which effectively filter out the AC voltage induced by RF excitation, thereby ensuring accurate measurement of DC thermal voltage. To demonstrate wireless operation, the distance between the transmitting and receiving antennas was set at 20 cm. The RF generator was tuned to operate between 35 and 45 MHz at 1W power. Given the absence of a tuned receiver circuit, the actual received power was significantly lower than the input power as detailed in the supporting information (**Figure S7**). **Figure 5c** represents the thermal image of RF heating at various input RF. We gave an input RF frequency from 35MHz to 45 MHz at 1W RF power. Frequency-dependent maximum temperature plot in the above frequency range is given in supporting information (**Figure S8**). A maximum temperature of $58.8^0$C was achieved at 35 MHz which can be attributed to the frequency resonance of the receiver setup. **Figure 5d** shows the frequency-dependent thermal voltage ($V_{TE}$) generated in the setup along with the temperature gradient ($\Delta T$). A maximum thermal voltage of -225mV with a high thermal gradient of 32K is observed at 35MHz. The negative potential is attributed to the n-type behavior of PbS. To verify that this voltage is coming from thermoelectricity, we recorded the RF input and output signals as shown in **Figure 5e**. It is observed that the output signal is showing some DC characteristics in the positive half cycle. To analyze this in detail, **Figure 5f** shows only the output signal. The output signal resembles that of a diode clipper where the positive part of the input signal is clipped by a voltage source. This voltage source can be thermal voltage generated in the device. To explain this, we created the equivalent circuit of the setup as shown in the inset of **Figure 5f**. The extra voltage source in this case is $V_{TE}$. The amplitude of the clipped signal is given by $V_{OP}=V_T + V_{TE}$, where $V_T$ is the turn-on voltage of the diode and $V_{TE}$ is the thermal voltage. The observed output voltage is 0.6 V and the turn-on voltage at 0.35 V, the thermal voltage $V_{TE}$ is calculated to be 0.25 V or 250 mV which is very close to the observed thermal voltage at 35 MHz. This finding justifies that the voltage obtained originates from thermoelectric voltage generated during RF heating. The RF thermoelectric power/ Seebeck coefficient is also calculated from the slope of Thermal voltage and Thermal gradient as shown in **Figure 5g**. The obtained Seebeck coefficient of -5.81mV/K is 14 times that of the maximum Seebeck coefficient observed for PbS (0.44mV/K) which is due to the high thermal gradient induced due to RF heating. Using the obtained Seebeck coefficient and electrical conductivity we calculated a power factor 151 mWm$^{-1}$K$^{-2}$. Finally, to calculate the power and efficiency, we measured the output voltage-current obtained from the device as shown in **Figure 5h**. Using these values, we calculated the frequency-dependent output power with a maximum of 225 µW. Taking the ratio of power received to power generated an overall efficiency of 15 % is obtained. Cascading such many devices along with a tuned resonant network matching circuit could increase the overall power output and efficiency of the device.

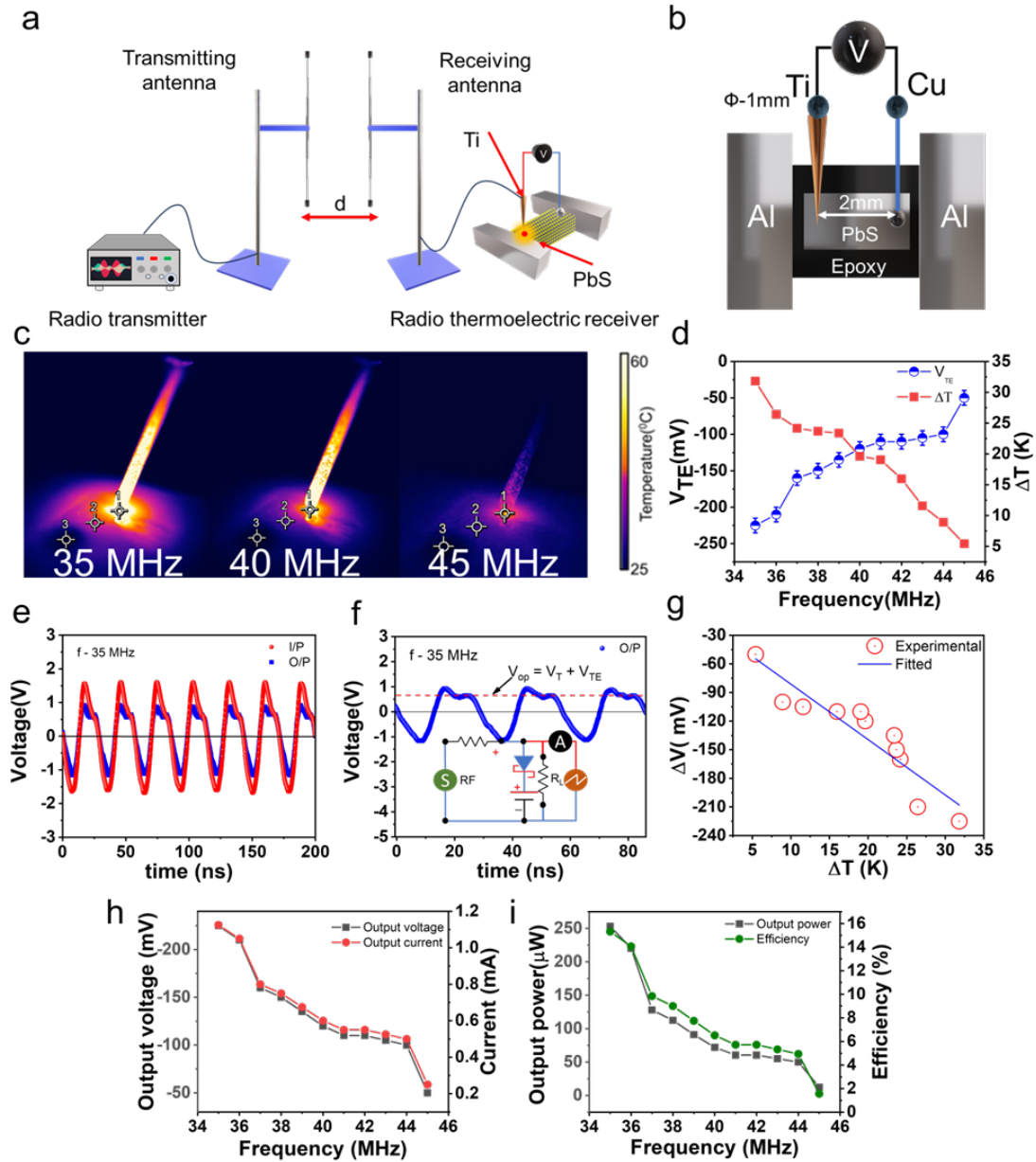

*Figure 5- (a) Schematic representation of radiofrequency heating assisted thermoelectric setup consisting of RF transmitter and RF thermoelectric setup, (b) Schematic representation of Ti/PbS contacts for RF heating and thermoelectric measurements (c)Thermal imaging photographs of Radiofrequency heating on PbS surface at 35MHz,40MHz and 45MHz (d) Frequency-dependent thermal voltage and thermal gradient plots of PbS,(e) Input and output voltage signal plots obtained from Ti/PbS junction, (f) Output voltage signal plots from Ti/PbS along with equivalent circuit diagram inset,(g) Thermal voltage and thermal gradient plots for calculation of average Seebeck coefficient, (h) Frequency-dependent output voltage-current plots and (i) Frequency-dependent output power and efficiency plots.*

## Molecular Dynamics Simulations and Evolution of Thermoelectricity in Local RF heated PbS

To understand the mechanism of the formation of a thermal gradient during RF heating in PbS, we performed molecular dynamics studies. **Figure 6a** represents the 3D representation of the equilibrated PbS structure for MD simulations. **Figure 6b** presents the calculated thermal conductivity as a function of temperature from 300 K to 550 K and the results are qualitatively consistent with the experimental results, which verifies the capability of the selected potential to capture the thermal transport properties of PbS. **Figure 6c** shows the schematic representation of heat flow initiation in PbS. The average temperature distributions of the system at 0 to 0.1 ns, 0.4 to 0.6 ns and 0.9 to 1 ns are presented in **Figure 6(d-f)**. At the beginning of the heating period when little heat is added to the system, the temperature distribution is uniform around 300 K (**Figure 6d**). When the system is heated at the highest rate (0.4 to 0.6 ns), a clear temperature gradient is observed where the region close to the heat source has a significantly higher temperature than the rest of the slab (**Figure 6e**). Near the end of the heating period (0.9 to 1 ns), the temperature difference still exists but is less apparent (**Figure 6f**). The temperature distributions revealed by MD simulations agree well with the experimental thermal images and provide more insights into the heat transport during one cycle of RF heating.

After confirming a thermal gradient generation during local RF heating in PbS, we next explain thermoelectric voltage generation. The RF heating at the metal-semiconductor (MS) interface originates from current due to Schottky emission when it is biased. When an RF signal is fed into the MS junction the diode conducts through thermionic emission of electrons that jump over the Schottky barrier. These hot electrons result in local heating of the junction. The whole process is repeated at high speed according to the input radio frequencies. This modifies the Schottky junction to act like a pulsed phonon source that can concentrate heat on specific regions of the material. Due to the pulsating nature of RF heating, we verified the mode of phonon transport in PbS by calculating the thermal penetration depth using the following equation.

$$\delta = \sqrt{k/C\pi f}. \qquad (1)$$

Where δ is the depth of thermal penetration, k is the thermal conductivity, C is the specific heat capacity and f is the frequency[34]. The δ-frequency plot is provided in the supporting information (**Figure S9**). Now to determine the mode of propagation of phonons we calculated the mean free path of phonons in PbS using kinetic model theory as depicted in the following equation:

$$k = \left(\frac{1}{3}\right) C v \Lambda. \qquad (2)$$

Where k is the thermal conductivity, C is the specific heat capacity, v is the phonon velocity and Λ is the mean free path.

From calculations, the mean free path of phonons in PbS is 1.4 nm. Since the mean free path of the phonon is less than the depth of penetration, the phonon transport will be in diffusive mode. As a result, we see localized heating in PbS. This in turn results in a large temperature gradient which causes the carriers/electrons to migrate towards a colder region, generating a large potential gradient / enhanced Seebeck voltage. The above concepts are depicted schematically in **Figure 6g.** From this, we conclude that localized RF heating can significantly enhance the thermoelectric effect of PbS. The same approach may be extended to other well-known thermoelectric materials which can be utilized for electromagnetic energy harvesting .

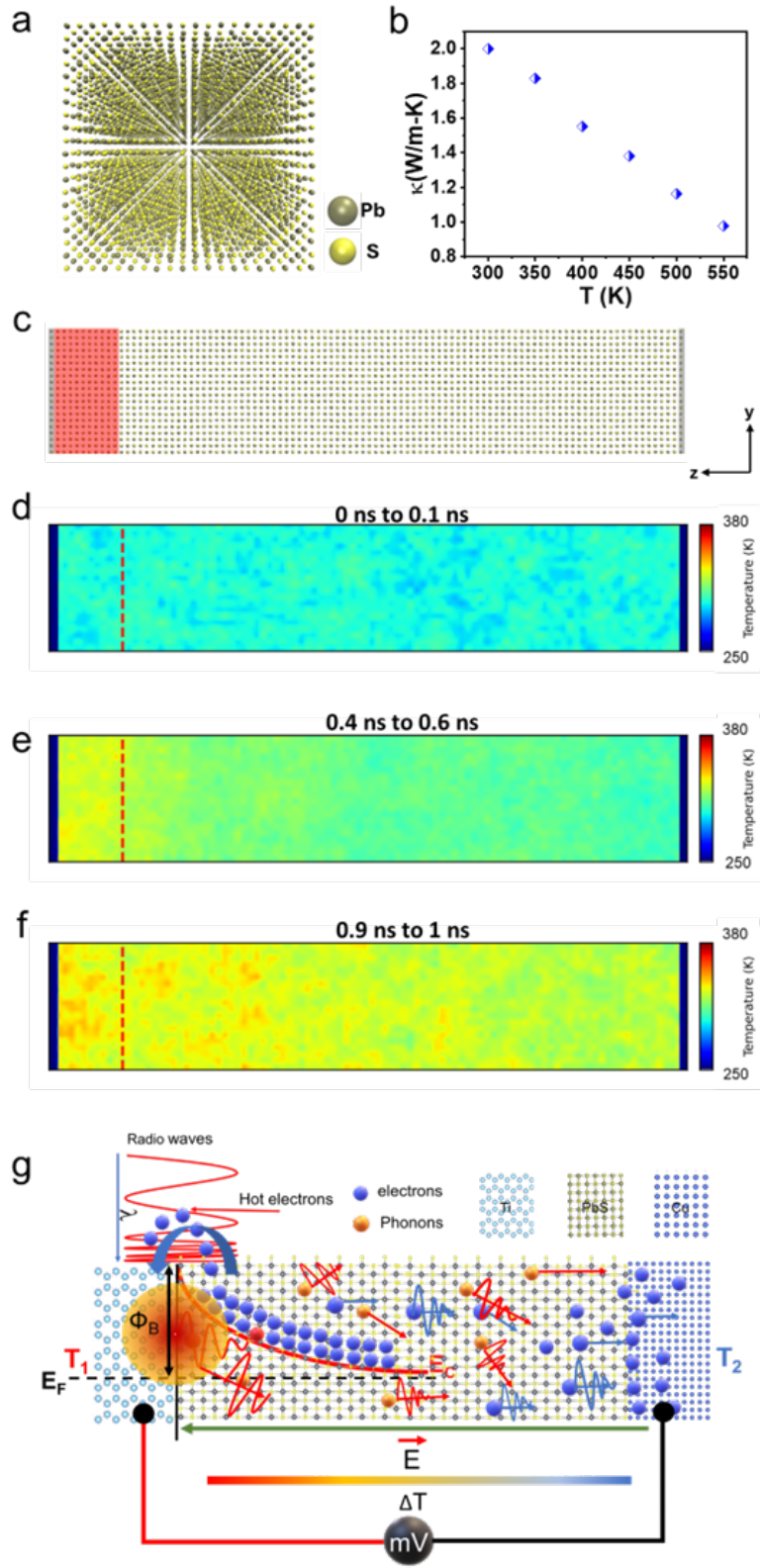

***Figure 6-*** *Molecular dynamics simulations of PbS. (a) A snapshot of the equilibrated cubic system, (b) Calculated thermal conductivity of PbS as a function of temperature, (c) A snapshot of the NEMD setup where heat is added to the region colored in red, (d-f) The temperature distributions in the system during stepwise heating that mimics sinusoidal heating by RF, (g)*

*Schematic representation of local RF heating induced thermoelectric effect in PbS using band diagram approach.*

## Conclusions

We demonstrated a novel approach to enhancing the thermoelectric performance of PbS by leveraging localized radiofrequency (RF) heating. Microscopic and spectroscopic analysis confirmed the structural, morphological, vibrational and chemical stability of polished PbS. Electrical characterization of the Ti/PbS junction revealed the formation of a Schottky junction with a low turn-on voltage of 0.3V and a low Schottky barrier height of 0.58 eV making it ideally suitable for RF detection applications. High-frequency AC electrical conductivity further validated the impact of the skin effect in lowering electrical conductivity at high frequencies, facilitating localized heating. Thermal imaging studies of Ti/PbS Schottky junction under RF exposure demonstrated maximum temperature of $58.7^0$C at 35MHz, generating a substantial high thermal gradient of 32K. This led to an enhanced Seebeck voltage of -5.8mV/K which is 14.6 times higher than the Seebeck coefficient in bulk PbS along with a significantly improved enhanced power factor of 151 $mWm^{-1}K^{-2}$. Angle-resolved photoelectron spectroscopy (ARPES) studies revealed that thermal broadening of localized heavy and light hole bands in PbS increases the density of heavy holes near the conduction band, contributing to an improved Seebeck coefficient and higher thermoelectric power factor at elevated temperatures. Molecular Dynamic studies further corroborated the observed thermal transport mechanism in RF-heated PbS. Our combined experimental and theoretical findings suggest that localized RF heating can enhance thermoelectric behavior in PbS, making it a promising candidate for stray electromagnetic energy harvesting.

## Materials and Methods

Phase analysis of PbS was done through X-ray diffraction (XRD) (Bruker D8 Advance with a Lynx eye detector using Cu-Kα radiation). The microstructure, elemental composition and morphology of polished PbS were done through Scanning electron microscopy (SEM) and energy dispersive spectroscopic studies (EDS) (ZEISS Sigma). The orientation of grain in PbS was verified using electron-backscattered electron diffraction (EBSD) studies. The vibrational nature of PbS was also analyzed using Raman Spectroscopy (WITec, UHTS 300VIS, Germany). The chemical oxidation states of PbS were analyzed using X-ray Photoelectron Spectroscopy (XPS) (Thermo Fisher Scientific makes Nexsa base). The voltage-current characteristics and dielectric measurements of Ti-PbS were done using a Source meter (Keithley 2450) and a Precision LCR meter (SM6026). AC electrical conductivity measurements were done using a vector network analyzer (VNA). The thermoelectric measurements of PbS were done in the Physical property measurement system (PPMS). RF heating and thermal imaging studies were done using a lab-made RF signal generator of 1W power and a Thermal imaging camera (Optris PI640). Thermal voltage measurements during RF heating were done using an analogue millivoltmeter.

## Photoemission Measurements

Photoemission measurements were performed at the undulator-based Angle-Resolved Photoelectron Spectroscopy beamline (ARPES, BL-10), Indus-2 synchrotron using a SPECS Phoibos 150 electron analyser. The single crystals of PbS were cleaved *insitu* at $5 \times 10^{-11}$ mbar to get an atomically clean surface. The base vacuum during the measurement was $7 \times 10^{-11}$ mbar. Synchrotron X-ray photoelectron spectroscopy (XPS) measurements were carried out using hν = 716 eV excitation energy at BL-10 with 0.3 eV energy resolution. High-resolution angle-resolved photoemission (ARPES) and valence band measurements were carried out at

300 K, 200 K, 100 K and 20 K using synchrotron radiation hv = 56 eV with 30 meV energy resolution and 0.1 deg angular resolution. The binding energy scale was calibrated with the Ag 3d lines and Ag Fermi edge following the standard procedure [35,36].

**Computational Methods-Molecular Dynamics**

To understand thermal transport induced by RF heating in PbS, we perform molecular dynamics simulations by Large-scale Atomic/Molecular Massively Parallel Simulator (LAMMPS) package[37]. We select the interatomic potential developed by Fan et al. [38], which is a combination of long-range Coulombic interactions and short-range Van der Waals interactions in the form of a Buckingham potential:

$$U_{ij} = \frac{q_i q_j}{r_{ij}} + A e^{-r_{ij}/\rho} - \frac{C}{r_{ij}^6}$$

where $U_{ij}$ Is the potential energy between atoms $i$ and $j$, $q_i$ and $q_j$ are atomic charges, $r_{ij}$ is the distance between $i$ and $j$, $A$, $\rho$ and $C$ are parameters. For interactions between Pb and S atoms, $A = 3.05 \times 10^6$ eV, $\rho = 0.173$ Å and $C = 154$ eV·Å$^6$. For interactions between S and S atoms, $A = 4.68 \times 10^6$ eV, $\rho = 0.374$ Å and $C = 120$ eV·Å$^6$. The short-range interactions between Pb and Pb atoms are ignored. The effective ion charges $q$ are $\pm 0.8e$.

To calculate the thermal conductivity of PbS, we create a cubic simulation cell with 20 atoms in each dimension. The system is minimized and equilibrated in an NPT ensemble for 0.1 ns. The lattice constant calculated from the equilibrated system at 300 K is 6 Å (**Figure 6a**), which is in good agreement with experimental value [39]. A production run in NVE for 20 ns where heat fluxes in x, y and z directions are recorded. The thermal conductivity is calculated by integrating the heat flux autocorrelation function using the Green-Kubo method[40]:

$$\kappa = \frac{V}{k_B T^2} \int_0^\infty \langle J(t_0) J(t_0 + t) \rangle \, dt$$

where $V$ is volume, $k_B$ is Boltzmann constant, $T$ is temperature, $J$ is heat flux, $t$ is correlation time and $\langle\ \rangle$ is the average overall time origins $t_0$. The production period is divided into 20 trajectories of 1 ns and the thermal conductivity is averaged over the results from 20 trajectories and three directions. The integrated thermal conductivity converges as $t$ approaches 25 ps (**Figure S10**) and the final value is determined at $t = 25$ ps.

To simulate the heat transfer under RF heating in PbS, we employ nonequilibrium molecular dynamics (NEMD) setup in a slab-shaped system with 20, 20 and 100 atoms in x, y, and z directions, respectively[41]. The system is divided into 100 layers in the z-direction, each containing 400 atoms. The left and rightmost layers are fixed as walls and periodic boundaries are applied only in x and y directions. The 10 layers next to the left wall are considered as an RF heat source (**Figure 6c**). Since LAMMPS only supports a constant heating rate, we apply stepwise heating to the heat source for 1 ns to mimic the sinusoidal heating by RF given in supporting information (**Figure S11**). The temperature distribution in the PbS slab is recorded to gain insights into the heat transfer during RF heating.